\documentclass[useAMS,usenatbib]{mnras}
\usepackage{graphicx,graphics}
\usepackage{amssymb}
\usepackage{amsmath}
\usepackage{epsfig}
\usepackage{float}
\usepackage{array}
\usepackage{comment}
\usepackage{multirow}
\usepackage{pdflscape}
\usepackage{caption}

\newcolumntype{L}{>{\centering\arraybackslash}m{2cm}}
\newcolumntype{K}{>{\centering\arraybackslash}m{1.5cm}}

\title[Electromagnetic transitions  of Sn IV]{Accurate estimations of electromagnetic transitions  of Sn IV for stellar and interstellar media}        
\author[S. Biswas et al.] {Swapan Biswas, Arghya Das, Anal Bhowmik\thanks{analbhowmik@phy.iitkgp.ernet.in}, Sonjoy Majumder\thanks{sonjoym@phy.iitkgp.ernet.in}
\\Department of Physics, Indian Institute of Technology Kharagpur, Kharagpur-721302, India }

\def\LaTeX{L\kern-.36em\raise.3ex\hbox{a}\kern-.15em  
T\kern-.1667em\lower.7ex\hbox{E}\kern-.125emX}

\begin{document}

\label{firstpage}

\maketitle

\begin{abstract}
Here we report on accurate ab initio calculations to study astrophysically important  electromagnetic transition parameters among different low-lying states of Sn IV. Our ab initio calculations are based on the sophisticated relativistic coupled-cluster theory, which almost exhausts many important electron correlations. To establish the accuracy of the calculations, we compare our results with the available experiments and estimates the transition amplitudes in length and velocity gauged forms. Most of these allowed and forbidden transition wavelengths lie in the infrared region,  and they can be observed in the different cool stellar and interstellar media. For the improvement of uncertainty, we use experimental energies to the estimations of the above  transition parameters. The presented data will be helpful to find  the abundances  of the  ion in different astrophysical and laboratory plasma.
\end{abstract}


\section{Introduction}
The spectrum of triply-ionized tin (Sn$^{3+}$) was first observed by Rao nearly a century ago \cite{Rao1926}. Accurate estimations of transition lines of Sn IV have astrophysical interest since their detection in the stellar medium \cite{O'Toole2004, Chayer2005}. Using the Hubble's Space
Telescope Imaging Spectrograph (STIS), \cite{O'Toole2004}  observed the presence of Sn IV in several sdB and sdOB stars (HZ44, HD4539, HD171858, HD185510, HD205805,  FFAqr, Feige48, Fiege65, CPD-64$^\circ$481, PG0342+026, PG1032+406,  PG1104+243, PG1219+534, HD149382, Fiege66, Ton S 227, PHL932, UVO1758+36) in the spectroscopic analysis  at temperature ranging from 22000 K to 40000 K. \cite{Chayer2005} detected the presence of Sn IV in the atmosphere of the cool  white dwarfs and observed five spectral lines of Sn IV in the ultraviolet (UV) spectra. \cite{Proffit2001} determined the tin abundance of the early B-main-sequence star AV304 in the small Magellanic cloud(SMC) by using archival STIS/HST G140M spectral data to measure the 1314.54 $\mathring{\text{A}}$ resonance line of Sn IV. 

Chemical evaluation models predict that about two third of tin-abundance in solar system is produced by the S-process, with most of the remainder due to the r-process \cite{Arlandini1999, Sneden1996}. \cite{Sofia1999} stated that tin is the first element to show a well-determined interstellar gas phase abundance, formed primarily by the main s-process.  Both Sn IV and Sn II can be observed in the different scenario of star evolution.  The detections of low ionized tin   in the spectra of Goddard High Resolution spectrograph (GHRS) \cite{Hobbs1993} have drawn the attention of precise experimental and/or theoretical works for high precision spectroscopy of these ions. The EUV photo absorption spectra  recorded, using the dual laser producing plasma (DLP) technique has promising application in lithography \cite{Lysaght2005}. \cite{Dunne1992} have given prediction of the strengths of EUV transition lines of Sn IV using similar procedure. The technological advancement of observational instruments even have resolution to identify hyperfine lines, which are signature of relativistic effect in atoms or ions. There have been few experimental works to measure the lifetimes of few energy states of Sn IV \cite{ Andersen1972, Kernahan1985, Pinnington1987}.  Theoretical calculations of energies of some low-lying states  of this ion are also available in literature \cite{Migdalek2000, Leszek2009, Ryabtsev2006, Ryabtsev2007}. To the best of our knowledge, there is no theoretical calculations of atomic properties for Sn IV which accounts correlations exhaustively.  Therefore, there is a requirement of the  highly correlated relativistic calculations of transition parameters of SN IV  in the determination of  its abundances at different stellar and inter-stellar media.

 In this paper, we use a non-linear  relativistic coupled cluster (RCC) theory \cite{ Lindgren1987, Bishop1991, Dixit2007a, Dutta2011, Dutta2012, Dutta2016, Bhowmik2017a, Bhowmik2017b, Bhowmik2018}   to calculate electromagnetically allowed and forbidden transition strengths and lifetimes of low lying energy states of Sn IV. The theory includes almost all kinds of many body correlations including core correlation, core polarization, pair correlation  \cite{ Dutta2016, Das2018} with high efficiency.  

In section 2, we brief about the RCC method  and provide the expressions of the spectroscopic parameters used in this paper. Section 3 compares our calculated results with the experimental and theoretical values extracted from literature, and provide precise spectroscopic data of transition lines of Sn IV, relevant for astrophysical or astronomical observations.

\section{theory}
The RCC theory is a well-established many body theory which accounts the correlation exhaustively. The correlated wave function for atoms or ions of $N$ electrons is generated using the single valence  Dirac-Fock reference state $|\Phi_v\rangle$. Single valence state means a single electron at '$v$'th valence orbital on top of the closed-shell Slater determinant state $|\Phi\rangle$. All the occupied (core) and unoccupied  orbitals of the closed-shell are generated at the Dirac Fock (DF) level under the potential of $ (N-1) $ electrons following Koopman's theorem (\cite{Szabo1996}). Using the RCC formalism, we can relate the correlated wave function corresponding to valence orbital '$v$' with its reference state $|\Phi_v\rangle$ as,
\begin{equation}
|\Psi_v\rangle = e^T(1+S_v)|\Phi_v\rangle
 \hspace {0.2cm} \text{where} , |\Phi_v\rangle = a^{\dagger}_v|\Phi\rangle.                   
\end{equation}
Here, the operator  $T$ considers the excitations from the core orbitals. $ S_v $ excites at least one electron from the valence orbital which gives rise to valence and core-valence excited configurations \cite{Lindgren1987}.     
  Corresponding amplitudes of these excitation operators are solved using matrix equations of the closed-shell and open-shell systems as discussed in \cite{Dutta2016}. In this present work the correlations are restricted to single, double and  triple excitations to calculate the correlated wavefunctions. Due to enormous computational effort and resource, less significant (much less than 1\%)   \cite{Lindgren1987}, \cite{Lindgren1985}, \cite{Haque1984}, \cite{Dutta2012}, \cite{Roy2014} quadrupole excitations are neglected in our calculations. The transition matrix element of an operator $ {\hat{O}} $ can be represented as follows
\begin{center}
\begin{eqnarray}
&&O_{ k\rightarrow i} =\frac{\langle{\Psi_k|\hat{O}|\Psi_i}\rangle}{\sqrt{\langle{\Psi_k|\Psi_k}\rangle{\langle{\Psi_i|\Psi_i}\rangle}}}, \nonumber \\
=&&\hspace{-0.7cm}\frac{\langle{\Phi_k|(1+S^{\dagger}_k)e^{T^{\dagger}}{\hat{O}}e^T(1+S_i)|\Phi_i}\rangle}{\sqrt {P_{kk} P_{ii}}}
\nonumber \\
=&&\hspace{-0.7cm}\frac{\big[\langle\Phi_k|\bar{O} + (\bar{O}S_{1i} + S_{1f}^{\dagger}\bar{O}) + (\bar{O}S_{2i} + S_{2f}^{\dagger}\bar{O}) + ...|\Phi_i\rangle\big]}{N},
\end{eqnarray}
\end{center}
where, the normalization factor $N$ contains overlap matrix $P_{ii}=\langle{\Phi_i|(1+S^{\dagger}_k)e^{T^{\dagger}}e^T(1+S_k)|\Phi_k}\rangle$ and core-correlated operator is defined as 
$\bar{O}$ =$ e^{T^\dagger}\hat{O}e^T$.  The subscripts '1' and '2' of open-shell cluster operator represents the single and double excitations. The matrix elements $(\bar{O}S_{1i} + S_{1f}^{\dagger}\bar{O})$ and $(\bar{O}S_{2i} + S_{2f}^{\dagger}\bar{O})$ yield the pair correlation and core polarization, respectively. The detail descriptions electric dipole ($E_1$), qudrupole ($E_2$) and magnetic dipole ($M_1$) matrix element with respect to orbitals are available in ref. \cite{Dutta2016}

   The transition probabilities, $A_{k\rightarrow i}$, of above electromagnetic multipole channels among the atomic states  $|\Psi_k\rangle$ and $|\Psi_i\rangle$ are given by
\begin{align}
g_k\lambda^3 A^{E_1}_{k\rightarrow i} &=2.0261 \times 10^{18}{|\langle{i|\hat{O}^{E_1}|k\rangle}|}^2 ,\\
g_k\lambda^5A^{E_2}_{k\rightarrow i} &= 1.1199 \times 10^{18}{|\langle{ i|\hat{O}^{E_2}|k\rangle}|}^2 ,\\  
g_k\lambda^3A^{M_1}_{k\rightarrow i} &= 2.697 \times 10^{13}{|\langle{i|\hat{O}^{E_2}|k}\rangle|}^2 .
\end{align}
Here  $ \lambda $ is the wave length of transition in $ \mathring{\text{A}} $ and degeneracy factor of the $k$-th state is $g_{k} = 2J_k + 1$ . The absorption oscillator strengths due to $E1$ transition  are given by \cite{Kelleher2008}
\begin{equation}
g_if_{ik} = 1.4991938\times 10^{-16}g_k\lambda^2 A^{E_1}_{k\rightarrow i}.
\end{equation}

The lifetime of a state $'k'  $  is calculated by considering all transition probabilities to the lower energy states $ 'i' $ and is given by 
\begin{equation}
\tau_k = \frac{1}{\sum_i {A_{k\rightarrow i}}}.
\end{equation}

\section{results and discussions}

The quality of   correlated wave functions  produced by  the RCC method  is  based on the generation of accurate  DF orbitals as explained in the previos section. To generate the precise wavefunctions of DF orbitals,  we use basis-set expansion approach  (\cite{Clementi1990}) in self-consistent field calculations. Here the radial part of each  basis considered  to have a Gaussian-type orbital (GTO)  having  even-tempered exponents \cite{Huzinaga1993} as basis parameters. The optimization of the parameters is performed \cite{Roy2015} such that the even-tempered basis based DF wave functions over radial extent mostly agree with the DF wave functions obtained using a sophisticated numerical approach, GRASP92 (\cite{Parpia2006}). For $s$, $p$, $d$, $f$, $g$ and $h$ symmetries, we have chosen 16, 15, 15, 11, 8 and 8 active  orbitals, respectively, for the RCC calculations out of 33, 30, 28, 25, 21 and 20 DF orbitals. This choice of active orbitals depend upon the convergence of the core correlation energy in the closed shell system (\cite{Dixit2008}).  The maximum difference between our RCC excitation energies  and NIST  \cite{NIST} results  occurs for $6s_{1/2}$ state, which is around  $0.45 \%$, and the average discrepancy between the two results is about 0.3 \%.

Table 1. shows  the electric dipole ($E1$)  transition amplitudes  in both length  and velocity gauge forms \cite{Grant2007, Johnson2006} along with the comparison  between our  calculated and NIST \cite{NIST} extracted  transition wavelengths. The correlation contributions to each of  the matrix  elements can be easily found from the difference between the DF and the RCC transition amplitudes  in both the gauge forms. Apart from $5s_{1/2}$ $\rightarrow $ $6p_{1/2,3/2}$ transtions, the average correlation contribution is about {5\%}. Including these two transitions the average correlation contribution becomes 23\%. It notifies  that the correlations are significant to those relatively weaker transitions.  The  table also shows good agreement between the length and velocity gauge results and  the average difference between them  is $ 3.2 \%$.  This agreement is one of   the uncertainty  estimations of our calculated wavefunctions. The consistent improvement of the ratio between these two gauged results further highlight the accuracy of our calculations.   This ratio is 1.06 and 1.01 for $5p \rightarrow 5d$  and $5p\rightarrow 6d$ transitions, respectively. Also, table shows 0.96 and 0.95 values of the ratio for $4f \rightarrow 5d$ and 
 $4f \rightarrow 6d$, respectively. Further accuracy can be analyzed by the consistency of the ratios, 20:10:2 (approximately) , of our calculated transition strengths among $^2P_{3/2}\rightarrow {^2}D_{3/2}$, $^2P_{1/2}\rightarrow {^2}D_{3/2}$ and $^2P_{3/2}\rightarrow {^2}D_{3/2}$, respectively \cite{cowan1981}. Most of the $E1$ transitions, presented here, fall in the  ultraviolet region of electromagnetic spectrum  apart from few and are  especially useful in space telescope based  astronomy (\cite{Goad2016}).   Some of the transitions,    $5s_{1/2}$ $\rightarrow $ $6p_{1/2,3/2}$,  $6s_{1/2}$ $\rightarrow $ $6p_{1/2}$, $4f_{5/2}$ $\rightarrow $ $6d_{3/2,5/2}$ and $4f_{7/2}$ $\rightarrow $ $6d_{5/2}$  belong to visible energy spectrum and can be used in laser spectroscopy.  

Near and Mid-infrared observations in Astronomy using space based telescope, Infrared Space Observatory \cite{Kessler1996} has opened different areas of astrophysical studies in cool region of space, like interstellar medium \cite{Feuchtgruber1997}, planetary nebulae \cite{Liu2001}. Table 2. presents such fine structure transition lines of Sn IV. Apart from their astrophysical importance these transitions present different correlation features of many-body formalism. Therefore, we present the forbidden electric quadrupole $(E2)$ and magnetic dipole $(M1)$ transitions of them. Like, electric dipole transitions, the accuracy of the calculations can be understood from the comparison of estimations using  formulations based on length and velocity gauges. In these transitions the difference is on average 5\% apart from $5g_{7/2}\rightarrow 5g_{9/2}$ $E2$ transition where we find large discrepancy. In the later case same discrepancy is observed even at the DF level using sophisticated numerical code GRASP92 \cite{Parpia2006}, which is not possible to mitigate by any correlated method.  Also,  our estimation for microwave transition among $4f$ multiplets show good agreement with relativistic configuration interaction calculations by \cite{Ding2012}. As expected, the correlation contributions to the fine structure $M1$ transitions are negligible and therefore for these transitions DF results are  very good approximation of the RCC values.

 In Table 3., we have listed oscillator strengths of $E1$ transitions, calculated with  their experimental transition wavelengths.  Due to the better stability of  the length gauge form of transition matrix elements compared to its velocity gauge form \cite{Grant2007}, we have used the former form of $E1$ transition amplitudes to calculate the oscillator strengths.  Our calculated results of oscillator strengths are compared with the previously reported theoretical [\cite{Leszek2009, Cheng1979, Migdalek2000, Migdalek1979}] and experimental data [\cite{Andersen1972, Pinnington1987}]. Table shows
excellent agreement of our results for $5s_{1/2}$ $\rightarrow $ $5p_{1/2, 3/2}$ transitions with the  calculations of  \cite{Leszek2009} using configuration interaction method based on DF wavefunctions generated with non-integer  (CIDF(q)) outermost core shell occupation number. They have shown that the contribution of fractional occupancy parameter at the DF level contributes around 20\% to the oscillator strengths. Their configuration space  is made up with the  single and double excitations as well as some  triple excitations from ground state  to a few low-lying states.  But, our present RCC calculations contain a larger active orbital space to exhaust to the correlation contributions.  The comparison with the old calculation by Cheng and Kim using  relativistic Hartree-Fock method \cite{Cheng1979} shows the correlation contributions in these evaluations. Migdalek and Baylis performed relativistic Hartree-Fock theory with core polarization using semi-empirically fitted polarization potential \cite{Migdalek1979} for few transitions.  Migdalek and Garmulewicz reported the oscillator strengths for a few transitions of Sn IV using two different methods \cite{Migdalek2000} and
they only differ by the treatment of valence-core exchange potential. The superiority of the present RCC method over all other many-body approaches discussed above that former  includes all leading order terms corresponding to core polarization, pair correlation, core correlation along with higher-order terms  for the transition matrix element calculations  \cite{Dixit2007a, Dixit2008, Dutta2016, Bhowmik2017b}. Our detail study of correlations show that the core polarization has  the dominant contribution in the total correlation in the presented $E1$ transition amplitudes. Therefore, our RCC results are in good agreement with all the core polarization augmented DF results. Some of the experimental data of oscillator strengths of $E1$ transitions of Sn IV were previously reported using Beam-foil technique \cite{Andersen1972, Pinnington1987} and they are seen to agree excellently with our RCC results. There is only disagreement seen for
$5p_{3/2}\rightarrow 5d_{5/2}$. According to the suggestion of  \cite{cowan1981}, the ratio of the $f-$ values for $5d_{5/2}\rightarrow 5p_{3/2}$, $5d_{3/2}\rightarrow 5p_{3/2}$ and $5d_{3/2}\rightarrow 5p_{3/2}$ transitions should be around 6:5:1, respectively. Our calculations show this ratio is 4.6:4.5:1, respectively, and therefore,  the Beam-foil experiment by \cite{Pinnington1987} underestimated the f-value.

Table 4. presents the transition rates of $E2$ and $M1$ transitions along with the corresponding experimental wavelengths. Most of the presented transitions are  either in the ultra-violet or in the infra-red (IR) regions. The transitions which fall in the ultraviolet region are very important, in general, in astronomical observation and plasma research (\cite{Saloman2004}; \cite{Morgan1995}; \cite{Fahy2007}; \cite{Morita2010}).  Recent study of forbidden transitions of monovalent atoms and ions by \cite{Safronova2017} suggests the  superior advantage of these transitions in numerical areas in physics and engineering, particularly precision measurement of time and fundamental constants. $5d_{3/2, 5/2} \rightarrow 6s_{1/2}$  transitions may have  applications in infrared laser spectroscopy and plasma research \cite{Thogersen1996}. Among all the presented $E2$ transitions, $5s_{1/2}$ $- $ $6d_{3/2, 5/2}$ transition matrix elements are maximally correlated, approximately 23\% and 27\%, respectively. All the other presented $E2$ transitions are less than 8\% correlated. However, the  $M1$ transitions, $5d_{3/2}$ $- $ $6d_{5/2}$ is abnormally correlated (around 82\%) due to the large pair correlation effect \cite{Dixit2007a, Dixit2008, Bhowmik2017b}. As seen from the table,  $5s_{1/2}$ $- $ $5d_{3/2, 5/2}$ transitions have stronger  probability (almost $10^5$s$^{-1}$) compare to other $E2$ transitions.  $M1$ transitions which have probabilities more than $10^{-3}$s$^{-1}$, are shown in the table. 


Since there is no metastable state of this ion, lifetime of the excited states are expected to be like neutral alkali atoms, of the order of nanosecond (ns). In Table 6,  we compare the present lifetimes for few low lying states of Sn IV with the theoretical as well as a few experimental results. The life times are calculated using present RCC amplitudes and the experimental wavelengths from the NIST \cite{NIST} to minimize uncertainity  due to the transition wavelengths.  Andr\'es-Garc\'ia et al. have used  the Griem semi-empirical approach using the COWAN computer code  \cite{Andres-Garcia2016}. The  other theoretical lifetimes calculated by Cheng and Kim \cite{Cheng1979} and Migdalek and Baylis \cite{Migdalek1979} are also presented in the table for comparison. The listed experimental  lifetimes of different excited states of Sn IV measured using Beam-foil spectroscopy \cite{Pinnington1987, Kernahan1985, Andersen1972}  are very close to our RCC lifetimes results apart from $5d_{5/2}$ due to discrepancy of the $5p_{3/2}\rightarrow 5d_{5/2}$ $E1$ transition as discussed in earlier paragraph.

The theoretical uncertainties in the calculated property parameters are evaluated by the quality of the wave functions at the levels of the DF. Also, we  consider  the uncertainty contributions from other correlation terms,  not considered in this work, and quantum electrodynamics effects. Later contributes  at most $\pm$2\%. Therefore, in this work, the maximum
uncertainties are $\pm$2.5\%  and $\pm$2.3\% for the allowed and forbidden transition amplitudes, respectively.

\section{conclusions}
This paper presents the transition amplitudes,  strengths and rates   of astrophysically important allowed  and forbidden transitions  for  the ion Sn IV using a highly correlated relativistic many-body approach. The transitions presented in this paper, are in ultraviolet, visible, infrared and microwave regions.  Our calculated estimations of most of  these transitions are in good  agreements  with the available experimental observations and theoretical calculations. The differences in results are explained and further justified with the help of estimations based on the length and velocity gauged forms.  The spectroscopic data of the present work will be useful in the estimations of abundance of the ion  in the different astronomical bodies,  astrophysical plasma, laboratory plasma, specially in stellar and interstellar media. Since some of the transitions are estimated first time in literature, to best of our knowledge, our calculated  data  may also help to the astronomer to discover the undiscovered lines in astronomical systems.

\begin{table*}
\centering
	\caption{The electric dipole transition matrix elements (in a.u.) at the DF \& the RCC levels of calculations for  both length  and velocity gauge forms. Also experimental transition wavelengths ($\lambda_{NIST}$) is compared with the same from our RCC calculations ($\lambda_{RCC}$). Wavelengths are in $\mathring{\text{A}}$ unit.}
	\begin{tabular}{c c c  c c c c c c c c c} 
		\hline 
		 & & & & &\multicolumn{3}{c}{ Length Gauge} &  & \multicolumn{3}{c}{ Velocity Gauge}   \\
		\cline{6-8}\cline{10-12}
		Transition & $J_l$ &$J_u$ & $ \lambda_ {RCC}$ & $ \lambda _{NIST}$   &DF && RCC &&DF & &RCC\\  
		\hline 
		$5s \longrightarrow  5p$&1/2 &1/2 &1440.8 &1437.5 &1.8671  &&1.5659 &&1.8226 &&1.5900 \\ 
		&1/2 &3/2 &1320.2 &1314.5 &2.6433  &&2.2253 &&2.5697 &&2.2536 \\
		$5s \longrightarrow  6p$&1/2 &1/2 &507.2 &505.4 &0.0812  &&0.1836 &&0.0914 &&0.1682 \\
	    &1/2 &3/2 &501.8 &499.9 &0.0329  &&0.1834 && 0.0512 &&0.1604 \\
		$6s \longrightarrow  6p$&1/2 &1/2 &4204.5 &4217.3 &3.6738  &&3.5135 &&3.6283 &&3.4928 \\
		&1/2 &3/2 &3859.4 &3862.2&5.1600  &&4.9406 &&5.0860 &&4.9058 \\
		$5p \longrightarrow  6s$&1/2 &1/2 &962.1 &956.3     &1.0166  &&1.0120 &&0.9921 &&0.9888 \\
		&3/2 &1/2 &1024.5 &1019.7 &1.5836  &&1.5645 &&1.5413 &&1.5181 \\
		$5p \longrightarrow  7s$&3/2 &1/2 &621.4 &619.0    &3.2234  &&3.3226 &&3.1607 &&3.2529 \\
		&1/2 &1/2 &597.8 &595.1     &0.3093  &&0.3139 &&0.2982 &&0.3073 \\
		$5p \longrightarrow  5d$&1/2 &3/2 &1041.4 &1044.5 &2.9490  &&2.6038 &&2.8498 &&2.5826 \\
	    &3/2 &3/2 &1114.9 &1120.7 &1.3294  &&1.2778 &&1.2702 &&1.2051 \\
		 &3/2 &5/2 & 1113.1 &1119.3 &4.0488  &&3.8480 &&3.8618 &&3.6150 \\
		 $5p \longrightarrow  6d$ &1/2 &3/2 &607.1 & 605.2 &0.5164 &&0.3803 &&0.4791 &&0.3815 \\
		&3/2 &3/2 &631.4 &630.0    &2.3877  &&2.3413 &&2.3499 &&2.3157 \\
		 &3/2 &5/2 &631.2 &628.7    &7.1195  &&7.0158 &&7.0037 &&6.9390 \\
		$ 6p \longrightarrow 7s$ &1/2 &1/2 &2528.7 &2514.8 & 2.0992 &&2.0612  &&2.0623  && 2.0253    \\
           &3/2 &1/2 &2672.4  & 2660.6 &3.2234 &&3.1575 &&3.1607 &&3.0923     \\
		$ 6p \longrightarrow  6d$&1/2 &3/2 &2703.6 &2706.7 &5.1306 &&4.9242 &&5.0383 &&4.8533 \\
		 &3/2 &3/2 &2868.6 & 2876.5 &2.3877 && 2.2918 && 2.3499 && 2.2660 \\
		 &3/2 &5/2 &2864.2 &2849.3 &7.1195  &&6.8656 &&7.0037 &&6.7880 \\
	   $ 5d \longrightarrow  6p$&3/2 &1/2 &3154.8 &3072.6 &3.0850  &&3.0584 &&2.9315 &&2.9094 \\
		 &3/2 &3/2 &2956.4 &2879.7 &1.3294  &&1.3231 &&1.2702 &&1.2634 \\
		  &5/2 &3/2 &2969.8 &2888.5 &4.0488  &&3.9850 &&3.8618 &&3.7922 \\
		 $5d \longrightarrow  4f$&3/2 &5/2 &2241.3 &2221.6 &5.5646  &&5.1718 &&5.7219 &&5.4041 \\ 	 
	    &5/2 &7/2 &2265.5 &2229.8 &6.6731  &&6.1593 &&6.9000 &&6.4668 \\
	    &5/2 &5/2 &2249.0 &2226.8 &1.4933  &&1.3840 &&1.5262 &&1.4361 \\	
	   $4f \longrightarrow  6d$&7/2 &5/2 &4090.7 &4020.9 &4.0987  &&3.8654 &&4.3571 &&4.0997 \\
	   &5/2 &3/2 &4154.7 &4085.4 &3.4912  &&3.3390 &&3.6742 &&3.4986 \\
	    &5/2 &5/2 &4145.6 &4030.7 &0.9218  &&0.8897 &&0.9572 &&0.9200 \\
	   $	4f \longrightarrow  5g$&7/2 &7/2 &2107.8 &2082.2 &1.3616  &&1.2583 &&1.3579 &&1.2613 \\
	   	&7/2 &9/2 &2107.8 &2082.3 &8.0560  &&7.4448 &&8.0346 &&7.4643 \\
		&5/2 &7/2 &2122.3 &2084.9 &7.0970 &&6.6454 &&7.0793 &&6.6629 \\	
		\hline	
	\end{tabular}

\end{table*}

\begin{table*}
\centering
\caption{Fine structure transition amplitudes (in a.u.) of Sn IV in DF and RCC level of calculation.}
\begin{tabular}{c c c c c c c c c c c c c } 
\hline 
	 & & &\multicolumn{3}{c}{ Length Gauge} &  & \multicolumn{3}{c}{ Velocity Gauge} & & &   \\
	\cline{4-6}\cline{8-10}
	Transition & $J_l$ &$J_u$    &$O_{DF}^{E2}$ && $O_{RCC}^{E2}$ &&$O_{DF}^{E2}$ & &$O_{RCC}^{E2}$ &$O_{DF}^{M1}$ &$O_{RCC}^{M1}$ &$O_{other}^{M1}$  \\  
	\hline 
$5p\longrightarrow 5p$ &1/2 &3/2 &6.5555  && 6.1617 &&6.2810 &&5.7857 &    1.1531 &1.1532 & \\
$6p\longrightarrow 6p$ &1/2 &3/2 &25.4611 &&24.5380  &&24.3819 &&23.4516 &1.1528  &1.1531 & \\
$5d\longrightarrow 5d $&3/2 &5/2 &8.1878 &&7.8169  &&7.9249 &&7.5636    &1.5490 &1.5488 &\\
$6d\longrightarrow 6d $&3/2 &5/2 &27.7864 &&26.9633 &&27.5002 &&26.6322  &1.5488 &1.5490 & \\
$4f\longrightarrow 4f $&5/2 &7/2 &8.7703 &&8.1111 &&7.0339 &&6.8043 &1.8515 &1.8511 &$1.8586^a$ \\
$5g\longrightarrow 5g$ & 7/2 &9/2 &21.4040 &&21.0925  &&98.3823 &&98.0656 &2.1082 &2.1082 & \\
\hline
a$\implies$\cite{Ding2012} \\	
\end{tabular}
\end{table*}
\begin{table*}
	\caption{Comparison of oscillator strengths ($f$) (in a.u.) of electric dipole transitions between the RCC and other endavours (experimental and theoretical). RCC results are obtained using experimental wavelength ($ \lambda _{NIST}$) as given here in $\mathring{\text{A}}$ unit. }
	\centering 
		\begin{tabular}{c r c c c c c    } 
		\hline \hline 
		Transition  & $J_l$ & $J_u$ & $ \lambda _{NIST}$ 
		&$ f_{RCC} $ & $ f_{Others} $\\  
		\hline 
		 $5s \longrightarrow  5p$ &1/2 &1/2  &1437.5  &0.259 &$0.307^{a}$,$0.258^{b}$, $0.350^{c}$, $0.243^{d}$ \\
		 & & & & & 0.263$^e$, 0.241$^f$, 0.240$^g$, $ 0.225^h $\\
	     & & & & &0.260$^i$, 0.240$^j$\\
		  & 1/2 &3/2  &1314.5   &0.572 &$0.671^{a}$,$0.567^{b}$, $0.764^{c}$, $0.538^{d}$ \\
		 & & & & &0.583$^e$, 0.534$^f$, 0.535$^g$, $ 0.500^h$\\
		 & & & & &	0.560$^i$, 0.640$^j$\\
		$ 5s \longrightarrow  6p $&1/2 &1/2  &505.4 &0.010 &- \\
		   &1/2 &3/2  & 499.9  &0.010 &- \\
		 $6s \longrightarrow  6p$ &1/2 &1/2  &4217.3  &0.445 &- \\
		  &1/2 & 3/2 &3862.2 &0.960 &- \\
		  $5p \longrightarrow  6s $&1/2 &1/2  &956.3  &0.163 &$0.165^d$, 0.161$^e$, 0.162$^f$, 0.159$^g$ \\
		  & & & & &	0.160$^j$\\
		    &3/2 &1/2  &1019.7  &0.182 &$0.185^d$, 0.182$^e$, 0.182$^f$, 0.180$^g$ \\
		   & & & & &	 0.180$^j$\\
		   $5p \longrightarrow  7s$ &1/2 &1/2  &595.1  &0.025 &- \\
		    &3/2 &1/2  &619.0  &1.354 &- \\
		 $6p \longrightarrow  7s $&1/2 &1/2 &2514.8 &0.257 &- \\
		 &3/2 &1/2 &2660.6 &0.285 &- \\  
		$5p \longrightarrow  5d$ &1/2 &3/2  &1044.5  &0.986 &$1.180^c$, $0.972^d$, 0.965$^e$, 0.963$^f$\\
	 & & & & &	 0.947$^g$, 0.943$^h$,  0.960$^j$\\	
	 &3/2 &3/2 &1120.7 &0.111 & $0.088^d$, 0.098$^e$, 0.097$^f$, 0.096$^g$ \\
	 & & & & &	 0.095$^h$\\
	  &3/2 &5/2  &1119.3  &1.005 &$1.069^c$, $0.885^d$, 0.881$^e$, 0.878$^f$, \\
	& & & & &	 0.866$^g$,  0.852$^h$, 0.630$^j$\\
	$5p \longrightarrow 6d$ &1/2 &3/2 &605.2 &0.036 &- \\
		  &3/2 &3/2  &630.0  &0.661 &- \\
		 &3/2 &5/2  &628.7  &5.945 &- \\
	$6p \longrightarrow  6d$ &1/2 &3/2 &2706.7 &1.360 &- \\
	&3/2 &3/2 &2876.5 & 0.139 & -\\
	&3/2 &5/2 &2849.3  &1.256 &- \\	 	 
	  $5d \longrightarrow  6p$ &3/2 &1/2 &3072.6  &0.231 &- \\
	   &3/2 &3/2 &2879.7  &0.046 &- \\
	  &5/2 &3/2  &2888.5  &0.278 &- \\
	  $5d \longrightarrow  4f$ &3/2 &5/2  &2221.6  &0.914 &$1.036^c$ \\
	   &5/2 &7/2 &2229.8  &0.861 &$0.977^c$ \\
	   &5/2 &5/2  &2226.8  &0.044 &- \\   
	  $4f \longrightarrow  6d$ &7/2 &5/2 &4020.9  &0.141 &- \\ 
	   &5/2 &3/2 &4085.4  &0.138 &- \\
	  &5/2 &5/2 &4030.7  &0.010 &- \\
	   $4f \longrightarrow  5g$&7/2 &7/2  &2082.2  &0.029&- \\ 
	  &7/2 &9/2  &2082.3  &1.011 &$1.100^c$ \\
	  &5/2 &7/2  &2084.9  &1.072 &$1.135^c$ \\
	\hline \hline
	\end{tabular}
		
\hspace{-2.5cm}	$a$ $\implies$ CIDF method with integer occupation number[\cite{Leszek2009}]. \\
\hspace{-1.65cm}	$b$ $\implies$CIDF(q) method with non-integer occupation number[\cite{Leszek2009}]. \\
\hspace{-5.1cm}	c $\implies$ Relativistic Hartree-Fock method [\cite{Cheng1979}]. \\
\hspace{-5.96cm}	$d$ $\implies $ DF+CP method[\cite{Migdalek2000}]. \\
\hspace{-2.30cm} $e$ $\implies $	DX+CP method with SCE model potential [\cite{Migdalek2000}].\\
\hspace{-1.66cm} $f$ $\implies $ DX+CP method with CAFEGE model potential [\cite{Migdalek2000}].\\
\hspace{-1.80cm}$g$ $\implies $ DX+CP method with HFEGE model potential [\cite{Migdalek2000}].\\
\hspace{-1.27cm}$h$ $\implies $	Semiempirical relativistic Hartree–Fock (Dirac–Fock) results  [\cite{Migdalek1979}] \\
\hspace{-7.10cm}	i $\implies $ Beam-foil technique  [\cite{Andersen1972}]\\
\hspace{-6.85cm}j	$\implies $Beam-foil technique  [\cite{Pinnington1987}]
\end{table*}

	\begin{table*}
	
	 \caption {DF and RCC transition rate (in s$^{-1}$) of $E2$ in length gauge ($A_{DF}^{E2} $ and $ A_{RCC}^{E2} $) and $M1$ ($ A_{DF}^{M1} $ and $ A_{RCC}^{M1} $) along with the experimental wavelengths ( $\lambda_{NIST} $) in  \AA. The notation $P(Q)$ for transition rates means $P\times10^{Q}$.} 
    \centering 
	\centering 
		\begin{tabular}{c c c c c c c c  c } 
	\hline 
			Transition &$J_l$ &$J_u$  & $ \lambda_{NIST} $ & $A_{DF}^{E2} $ & $ A_{RCC}^{E2} $ & $ A_{DF}^{M1} $ & $ A_{RCC}^{M1} $  \\ 
			\hline\\
			$5s-5d$ &1/2 &3/2  & 604.9 &1.0598 (+05) &9.2311(+04) & &  & \\
			  &1/2 &5/2  &604.6 &1.0563(+05) &9.2260(+04) \\
			$5s-6d $&1/2 &3/2  &425.9 &1.6629(+04) &1.2858(+04) & &  & \\
			  &1/2 &5/2 &425.3 &1.7794(+04) &1.2912(+04) & & \\
			  $5p-6p$ &1/2 &3/2  &766.5 &1.4222(+04) &1.3055(+04) &1.6191(+01) &1.3863(+01) \\
			 	 &3/2 &1/2  &821.2 &2.8691(+04) &2.6272(+04)&2.9772(+01) &3.4242(+01) \\
			 	&3/2 &3/2  &806.7 &1.3823(+04) &1.2681(+04) & & \\
			 	 
			 $ 5p-4f$&3/2 &7/2 &745.2 &9.5768(+04) &8.4420(+04) & & \\
			  &3/2 &5/2  &744.9 &2.1323(+04) &1.8862(+04) &$2.1018(-03)$ &$2.5918(-03)$  \\
			 &1/2 &5/2  &710.5 &8.5436(+04) &7.5457(+04) & & \\
		
			 $6p-4f$ &3/2 &7/2 &9778.1 &1.2745$(+00)$ & 1.1060$(+00)$ & & \\ 
		     &3/2 & 5/2 & 9720.6 &2.9366$(-01)$ &2.6202$(-01)$&  & \\ 
			
			$5d-6s$ &3/2 &1/2  &11319.8 &5.4750$(-01)$ &5.1505$(-01)$ & & \\
			&5/2 &1/2  &11457.4 &7.9011$(-01)$ &7.3143$(-01)$ & & \\
			$5d-7s$ &3/2 &1/2  &1382.9 &1.0096(+03) &9.7368(+02) & & \\
			
			 &5/2 &1/2  &1384.9 &1.6291(+03) &1.4869(+03) & & \\
			$5d-6d $&3/2 &3/2  &1439.0 &2.7130(+03) &2.5876(+03) & & \\
			 &3/2 &5/2  &1432.2 &7.7068(+02) &7.5189(+02) &1.7655$(-01)$ &6.1292$(-03)$ \\
			 &5/2 &5/2 &1434.4 &3.2066$(+03)$ &3.0232$(+03)$ & & \\
			$5d-5g$ &3/2 &7/2  &1075.5 &4.2270(+04) &3.9004(+04) & & \\
		
			 &5/2 &7/2  &1076.8 &4.8024(+03) &4.3446(+03)  & & \\
			 &5/2 &9/2  &1076.8 &4.8033(+04) &4.3457(+04) & &   \\
			$6d-5g$ &5/2 &7/2 &4318.7 & 4.7514$(+01)$&4.5416$(+01)$ & & \\
			 &5/2 &9/2 &4318.8 &4.7502$(+02)$ & 4.5402$(+02)$ & & \\
		
		\hline  \\
		
		\end{tabular}
		
	   \end{table*}

\begin{table*}
\begin{center}
	\caption{Lifetimes  in ns of few low-lying states. }
	\centering 
	\begin{tabular}{c c c c c  } 
		\hline\hline 
		Level &present work &other work(experiment) &other work(theory) \\ 
		\hline
		$5p_{1/2}$ &1.20 & 1.29
		$\pm0.20^a$, 0.73 $\pm0.40^b$& $1.03^d$,  $0.95^e$, $0.89^f$ \\
		$5p_{3/2}$ &0.90 & 0.81$ \pm0.15^a $, 0.93$\pm0.23^b$,   
		&$0.79^d$,  $0.74^e$, $0.68^f$ \\
		&  & 1.00$\pm0.10^c$  & \\
		$5d_{3/2}$ &0.28 &  0.34$ \pm0.04^a $, 0.35$\pm0.03^b$ &$0.23^d$,  $0.26^e$, $0.33^f$ \\
		$5d_{5/2}$ &0.28 & 0.45$\pm0.05^a$, 0.41$\pm0.03^b$, &$0.30^d$, $0.29^e$, $0.35^f$ \\
		& & 0.52$\pm0.08^c$ & \\
		\hline	 \\
		\end{tabular}
\end{center}
		a$\implies$ \cite{Pinnington1987}, 
b$\implies$ \cite{Kernahan1985},
c$\implies$ \cite{Andersen1972} \\
 d$\implies$ \cite{Andres-Garcia2016} 
e$\implies$ [\cite{Cheng1979}],
f$\implies$[\cite{Migdalek1979}] \\

 \end{table*}

  
\label{lastpage}
\end{document}